# Shadow Detection and Geo-tagged Image Information Based Strategic Infrastructure Characterization

**E. MD. Abdelrahim**
Department of Mathematics
Computer Science Division, Faculty of Science, Tanta University, Egypt
elsaid_abdelrahim@yahoo.com

**Romany F. Mansour**
Department of Mathematics
Faculty of Science, N.V., Assiut University, Egypt
romanyf@aun.edu.eg

**Abdulsamad Al-Marghilnai**
College of Computer Science & Information, Northern Border University, Arar, Saudi Arabia
srd.nbu@gmail.com

**ABSTRACT**
Exploring the significance of geo-tagged information of a satellite image and associated shadow information in the image to enable man-made infrastructure dimensional characterization, in this paper a novel approach has been developed to estimate the height of a man-made structure accurately. Our proposed method applies shadow detection and shadow feature along with geo-tagged information retrieved through Google Earth tools, to estimate the structural dimension of the man-made infrastructure. To characterize the dimensional features of the man-made structure, we have considered Google Earth satellite images with precise and crisp edge as well as shadow length information, which have been obtained using freely available Google Earth information retrieval tool. Applying our proposed mathematical formulation, the height of man-made structures as well as neighbouring structures have been estimated which can be used for certain strategic decision purposes. The proposed method can be applicable for professionals exploring dimensional characterization of relatively low height constructions as well as defence purposes. Our proposed model has outperformed majority of existing approaches, particularly in terms of accuracy of height estimation of man-made structures.

*Keywords—Shadow Information; Satellite image; Height Estimation; Dimensional Characterization; Geo-tagged information*

## 1. INTRODUCTION

The high pace rise in technologies and associated applications have motivated global academia-industries to explore and enable optimal solutions to make human life more constructive, comfortable and productive. No doubt, to enable human life, whether being a regional or international across human existence horizon, technologies have made their un-substitutable place. Being human by nature, laughable but highly sensitive matter, efforts have been made to strengthen strategic infrastructures, strategic solutions including the weapon systems to ensure safe home nations and to counter opponent. There should not be a second thought that that almost all countries are making efforts to achieves dominating technologies and systems to counter offensive entities for national security. In recent years, the research domain which has grabbed the attention of global academia-industries is Autonomous Unmanned Vehicles (AUVs) and defense systems. One of the dominating technical needs for such systems are to characterize natural or even human made infrastructures to make proper decision. Here, the infrastructure can be the height of building, dimension of a building and its architecture etc. Extracting such critical information might help agencies or even AUVs to make proper decision. Interestingly, the purpose could be to identify strategic infrastructures such as height of a building or construction, dimension of a construction, nature of man-made construction and its relevance to certain reference.

A well known feature called shadow can be of paramount significance to characterize strategic infrastructure under sun-light presence. Exploring shadow and associated features including dimensions (length, width etc) can be helpful for strategic infrastructure identification. A shadow is formed when certain object comes in the way of a light source. It is cast by certain occluding object, or the object itself may be shaded. Because of the dissimilarity in between the intensity of the light reaching a shaded region and an unswervingly lit space, shadows can be characterized through related gradient of the brightness. On the contrary, non-shadow spaces can be illuminated by means of either the direct medium or the diffuse light medium. Here, direct medium represents sunlight and flashlight, while diffuse medium signifies the skylight, fluorescent, or incandescent source of light. The variations in between the shadow and the non-shadow regions do signify not only the brightness distinction, but a color one too. For instance, in case of outdoor condition, the lit region used to be illuminated by natural lights (i.e., sunlight, skylight), while the shadow area illuminated by skylight can generate or cast a bluish color. Such shadow characteristics, in conjunction with the fact that there can be certain color in both the lit as well as shaded objects, enables it stochastic in numerous vision based computations such as tracking [1], object detection and identification [2], and white-balancing [3] etc. With such immense utilities, the issue of shadow detection has gained global attention across industries to enhance the performance of computer vision techniques [4-8].

In practice, shadow detection suffers from numerous issues, primarily due to change in the shadow objects dimension (shape, orientation, and the physical properties) and the source of light. In comparison to the inclusive appearance paradigm, sun orientation and associated shadows are the sophisticated characteristics, yet are having potential towards characterizing certain outdoor infrastructure or scene appearance. Shadow maps can be applied for different real time textural mapping purposes, particularly for 3D models [9]. As inclusive model, the consideration of shadow and its direction can be vital towards outdoor photometric stereo purposes, which has gained attention to enable significant (fine-grained) geometric information better as compared to the multi-view stereos [10]. Estimating the direction and position of sun is also significant to correct timestamps in Internet images [11, 12], where there can be time zone differences and imprecise setting of the camera clocks causing the image metadata error prone and unreliable.

With the goal to extract information such as building heights, etc using shadow details a number of studies have been done. Numerous studies have been done to estimate the height of a building using certain shadow detection technique, whoever authors have emphasized on employing those dimensions for energy consumption estimation [12-20]. However, field measure based approach is highly resource and time consuming, likelihood of being biased or manipulated, etc. Particularly for defence systems in major cases its infeasible and risky too to physical perform infrastructure characterization by field view. No doubt, it seems ineffective in present day scenarios where technologies as well as targeted monitoring systems have increased significantly. To estimate the infrastructure characterization, especially for dimensional estimation different efforts have been made using different datasets such as Digital Terrain Model (DTM), Digital Surface Model (DSM) using sensor data [21]. However reconstructing it (i.e., 3D model of physical infrastructure) for decision purpose is highly intricate task. In [22] authors applied SAR airborne cameras to obtain high-resolution images and performed stereoscopic analysis to extract 3D building. Similar effort was made in [23] where authors used DTM and SAR sensor data to estimate building height. Authors applied Probabilistic Criterion Optimization model [23]. LiDAR based images were used to perform energy consumption estimation oriented building dimensional characterization [24]. Recently, authors applied images retrieved from interferometric synthetic aperture radar (InSAR) and aerialorthophoto to calculate the height of a building having place or flat roof structure [25]. These literatures state that employing images and associated shadow information dimensional characterization of the physical man made infrastructures can be done effectively.

A single high-resolution image based building detection model was developed using shadow detection and Markov random field (MRF)-based region growing segmentation technique [26]. Authors in [27] emphasized on employing GPS in radar networks employing the effect of Forward Scattering (FS) of electromagnetic waves for detecting, locating and classifying the objects through their GPS radio shadows. Using singular satellite image, a polygonal shape roof building height estimation model was developed in [28]. Authors [28] applied image features such as lines, line-intersections and employing graph-based approach established a roof-concept region for building height estimation. The concept region estimation applied shadow and geometry information, which was further processed with Fuzzy-logic algorithm to perform height estimation. Similar effort was made in [29], where authors used satellite images to detect buildings and its height. They used key information such as height, color, and shape to detect man made structure. They applied energy minimization (within a definite area) model to estimate the height of the object. Adjusting the shadows length and building height, authors [30, 31] developed a building height estimation model. Using roof size details by Google Earth, author [32] estimated four corner points of the rectangular facade of the building and then employing intrinsic parameters of the camera, they estimated the building height. Recently a model using relationship between shadow and building height was proposed by Shao et al [33]; however they found that over-simplification of such relationship might introduce significantly large error, even of 15 m difference from the actual building [33]. Author [34] derived a model to estimate building heights based on the length of the shadow, solar elevation and azimuth, satellite elevation and its azimuth. Similar effort was made in [35, 36] where authors applied sine-cosine theorem employing shadow length, solar elevation, and satellite elevation. Additionally, they applied the difference between solar azimuth and satellite azimuth.

Recently, a conventional but technically enriched technique named Google Earth has came into existence which has proved its excellence to track a number of activities comprising the strategic governmental or non-governmental infrastructures, proliferation of weapons for certain probable mass destruction etc. So far, its quantitative potential was confined till estimating the horizontal distances between different infrastructures or locations. However, exploring inside, it can be found that Google has prohibited providing any significant information to permit users to employ shadow dimensions for estimating heights of an object. It doesn't even provide any time stamp of the images retrieved through Google Earth applications. One prime reason could be the nature of the images exhibiting a mosaic of satellite and other related images, retrieved on different dates, locations and times, and amalgamated together to generate an approximates presentation. In recent versions of the Google Earth, the facility of displaying date information of the particular snaps is provided [37]. Interestingly, with the provided time information (i.e., date, month and year), employing compass heading of the shadow itself the reliable regional "time", analogous to a sun dial can be employed to estimate the time at certain instant and position.

In this paper robust shadow detection, shadow parameter estimation and geo-tagged information based strategic infrastructure characterization model has been developed. Our proposed model employs a novel shadow detection model along with geo-tagged information retrieved through Google Earth to estimate the infrastructure dimensional features. Being shadow detection based paradigm, our proposed model emphasizes on estimating the height of a natural or man-made infrastructure which can be applied for different strategic decision purposes.

The other sections of the presented paper are divided as follows. Section 2 discusses the proposed shadow detection and shadow length estimation model, which is followed by Geo-Tagged information retrieval (using Geometry and Information Retrieval Tool (GIRT)) from Google Earth. The discussion of the results obtained and their respective significances are presented in Section 3. Section 4 presents the conclusion and future scopes. References used in this research are presented at the end of the manuscript.

## 2. SHADOW DETECTION AND GEO-TAGGED INFORMATION BASED STRATEGIC INFRASTRUCTURE CHARACTERIZATION

In this paper, a multi-tech model consisting enhanced shadow detection and Geo-tagged information extraction models such as Google Earth have been applied to perform strategic infrastructure characterization or dimensional estimation model. Considering a free ware tool Google Earth doesn't provide key information such as building or geographical (including natural or man-made infrastructures) width, height etc, in this work, it has been applied to estimate shadow length and solar parameters. The well known Google Earth application, Geometry and Information Retrieval Tool (GIRT) has also been taken into consideration to estimate shadow height automatically. It strengthens the proposed model to reduce infrastructural length error components. Furthermore, to retrieve the geo-tagged information of the image, free version of the Google Earth has been applied. The overall proposed model and schematic implementation is presented as follows:

As depicted in Fig. 1, the overall proposed model comprises four phases. These are:

1. *Satellite Image retrieval,*
2. *Shadow detection and shadow length estimation,*
3. *Geo-tagged information retrieval using Geometry and Information Retrieval Tool (GIRT, Google Earth),*
4. *Target identification and distinction from similar infra-presence.*

As stated above, in the proposed model, at first the outdoor satellite image is obtained which is relevant for any defence or strategic purposes. Here, with ease of availability Google Earth data has been taken into consideration. Once retrieving the image for further analysis (i.e., dimensional estimation), shadow detection model has been executed. In addition, as a parallel approach, Google Earth's GIRT has been applied that estimates the key features of the shadow. Once retrieving the length of the shadow, to distinguish target construction or any natural/man-made infrastructure in the same satellite image, the ratio of the targeted infrastructure height to the length of its shadow has been estimated. It should be noted that GIRT can effectively enable height estimation. In our work, corner shadow length ratio (CLSR) approach [38] has been taken into consideration.

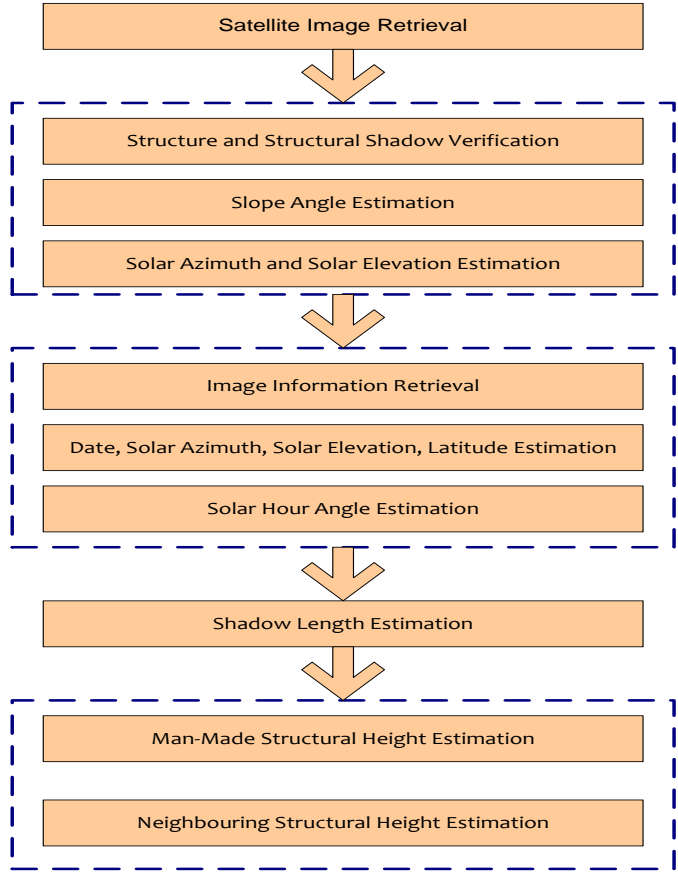

**Fig.1.** Shadow Detection and Geo-tagged information based structural height estimation

The discussion of the proposed model is presented in following sub-sections.

## 2.1 Image Validation

To ensure optimal or accurate shadow detection, shadow length estimation and (target) infrastructure characterization, assuring image with all expected key information is must. In this work, at first image validation has been done where the presence of certain structure is affirmed along with its shadow presence. Further, slope test, solar elevation and solar azimuth verification has been done using Google Earth GIRT tool. The discussion of the associated functions is given as follows:

*1) Structure and Structural Shadow Verification*

Before putting image for dimensional estimation and characterization, in the proposed model the selected satellite image has been tested for the two minimal requirements. These two criterion are; in the satellite image minimal one vertical edge should be distinctly visible so as to enable accurate shadow length estimation using proposed shadow detection and assessment model that applied Google GIRT tool. The other dominant criteria are that the shadow must be clear enough to avoid ambiguities.

## 2.2 Slope Angle Estimation

This is the matter of fact that the terrain slope angle is one of the key factors that influences length of the structural shadow. The details for slope detection and derivations can be found in [38]. With the available solar elevation ($h_s$) and slope angle ($\theta$), the relative error $\delta_H$ has been estimated using following equation:

$$\delta_H = \begin{cases} tanh_s \left|1 - \dfrac{\sin h_s}{\sin(h_s + \theta)}\right| Positive\ slope \\ tanh_s \left|1 - \dfrac{\sin h_s}{\sin(h_s - \theta)}\right| Negative\ slope \end{cases} \quad (1)$$

Thus, the relative error ($\delta_H$) rate with the highest positive and negative slopes and distinct solar elevations (angles) have been estimated and presented in Table 1 and Table 2.

**TABLE 1** RELATIVE ERROR WITH SOLAR ELEVATION ANGLE AND POSITIVE SLOPE

|  | $20^0$ | $30^0$ | $40^0$ | $50^0$ | $60^0$ | $70^0$ |
|---|---|---|---|---|---|---|
| $1.0^0$ | 0.02 | 0.02 | 0.02 | 0.02 | 0.02 | 0.02 |
| $1.5^0$ | 0.02 | 0.02 | 0.03 | 0.03 | 0.03 | 0.03 |
| $2.0^0$ | 0.03 | 0.03 | 0.03 | 0.03 | 0.03 | 0.03 |
| $2.5^0$ | 0.04 | 0.04 | 0.04 | 0.04 | 0.04 | 0.04 |
| $3.0^0$ | 0.05 | 0.05 | 0.05 | 0.05 | 0.05 | 0.05 |
| $3.5^0$ | 0.05 | 0.05 | 0.06 | 0.06 | 0.06 | 0.05 |

**TABLE 2** RELATIVE ERROR WITH SOLAR ELEVATION ANGLE AND NEGATIVE SLOPE

|  | $20^0$ | $30^0$ | $40^0$ | $50^0$ | $60^0$ | $70^0$ |
|---|---|---|---|---|---|---|
| $1.0^0$ | 0.02 | 0.02 | 0.02 | 0.02 | 0.02 | 0.02 |
| $1.5^0$ | 0.03 | 0.03 | 0.03 | 0.03 | 0.03 | 0.03 |
| $2.0^0$ | 0.04 | 0.04 | 0.04 | 0.04 | 0.04 | 0.04 |
| $2.5^0$ | 0.05 | 0.05 | 0.05 | 0.05 | 0.05 | 0.05 |
| $3.0^0$ | 0.06 | 0.06 | 0.06 | 0.06 | 0.06 | 0.06 |

Observing above mentioned error details, it can be found that the sola elevation and the slope are the key parameters affecting relative error in structural dimensional characterization using GIRT. Explores field measurement outcomes, particularly for slopes [39], it can be found that the lower relative error in dimensional characterization can be obtained only when the positive and the negative slopes are less than 3.00 and 2.500, respectively. It also affirms the generally believed hypothesis that the terrain around the man-made structures is typically flat.

Considering above revelations, it can be stated that slope angle does have significant impact in deciding whether the reason under target belongs to the rural or urban. It can be helpful for UAVs to make optimal decision towards intended purposes. In this empirical study, DEM has been taken into consideration to estimate slope. The more details about DEM based slope estimation can be found in [38].

### 2.3 Solar Elevation and Solar Azimuth Estimation

The overall proposed method of man-made structural dimension estimation is based on the hypothesis that the ultimate ratio of a height of the structure such as building's height to the length of the shadow remains constant. One of the key requirements needed inevitably is that the solar parameters such as solar azimuth and solar elevation in the remote sensing or satellite image must be close enough or equivalent. The fundamental parameters of the sun are depicted in following figure (Fig. 2). Such predominant needs can be easily fulfilled by the remote sensing (i.e., satellite) images retrieved through Google Earth [38]. These facts affirm that the distinct or separate assessment of the solar azimuth and elevation is not must to be carried out.

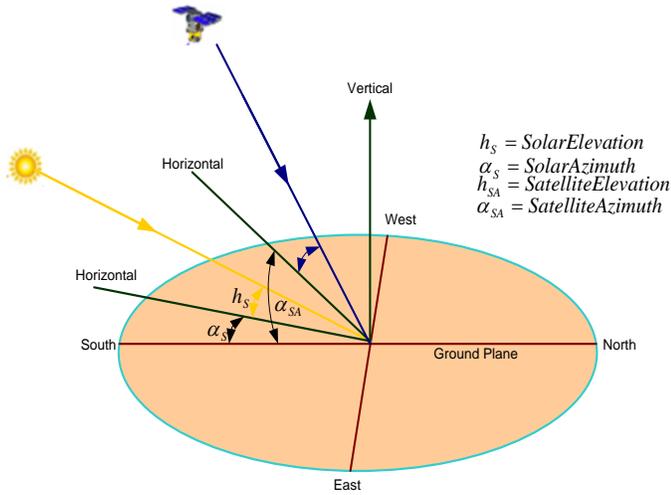

**Fig.2. Fundamental parameters of the sun and the satellite**

## 3. SUN-SHADOW PARAMETER ESTIMATION FOR MAN-MADE INFRASTRUCTURE CHARACTERIZATION

Once assessing the images for its appropriateness, the infrastructure characterization, especially in terms of height of the man-made infrastructure has been done. On the basis of the information retrieved straightforwardly from the remote sensing images of the satellite images obtained from Google Earth application. Here, in the proposed work, height estimation of the man-made infrastructure has been obtained as per celestial geometry concept. A brief discussion of the further dimensional (man-made infrastructure height) characterization is presented as follows:

### 3.1 Remote Sensing Image Information Retrieval

In the proposed work, the overall key information obtained from the remote sensing images through Google Earth is split into two sections. The first section comprises significant information such as acquisition time at instant, present latitude and altitude information. This information has been obtained directly from the single remote sensing image. Now, the remaining information comprises the key information such as shadow length $L_{A1A2}$, solar azimuth $\alpha_S$, satellite azimuth $\alpha_{SA}$ and vertical edge length $L_{A2B}$. These all key information has been estimated using GIRT tools of the Google Earth tool.

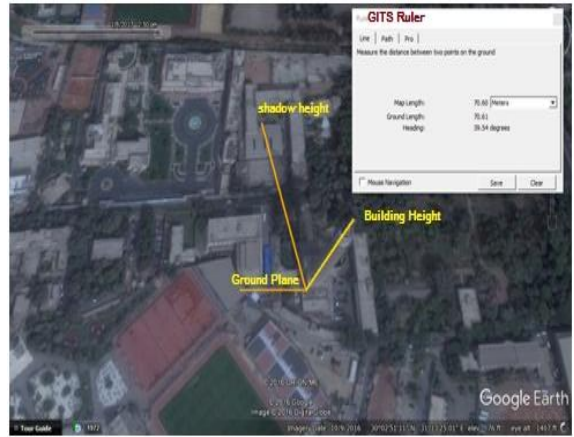

**Fig.3 Google Information Retrieval System (GITS) based geo-tagged information retrieval**

### 3.2 Solar Declination

The angle in between the rays of sun and the plane of earth equator is stated as the "Solar declination($\delta$). It is one of the fundamental parameters characterizing the location or the position of the sun at certain instant of time. A number of approaches have been proposed to estimate solar declination [40–45]. There are the techniques [46] that can ensure error equal or less than 0.010. Unfortunately, to achieve higher accuracy these approaches turn into highly complicate and to design solar position tracking systems. To alleviate such issues and maintaining optimal balance between accuracy and ease of implementation in this paper the following equation (2) has been used.

$$\delta = 0.3738 + 38.2567 \sin w + 0.1149 \sin 2w \\ - 0.1712 \sin 3w \\ - 0.7580 \cos w \\ + 0.3656 \cos 2w \\ + 0.0201 \cos 3w \tag{2}$$

where

$$w = \frac{360(n - n_0 - 0.5)}{365.2422} \tag{3}$$

The above derived equation [27] can be applied to estimate the solar declination for certain time or day [27].

Here, $n_0$ represents the spring-equinox time, $n_0$ presented in days initiating from the start of the year. Mathematically, $n_0$ is obtained as:

$$n_0 = 78.801 + 0.2422(YEAR - 1969) \\ - Int(0.25(YEAR - 1969)) \tag{4}$$

Here, it should be noted that Int() function represents the integer function.

As a strength or weakness considered per individual, in this study the images under study were taken during the solar declination from 11 AM to 2 PM under fair sun light. It enables proper image conditioning to perform height estimation. Considering the time of image retrieval, the respective solar declaration at different points has been obtained (Table 3).

**TABLE 3  SOLAR DECLINATION AT DIFFERENT TIME INSTANTS**

| Time\Date | 1.15 | 3.15 | 5.15 | 7.15 | 9.15 | 11.15 |
|---|---|---|---|---|---|---|
| 9:00 AM | -21.11 | -2.21 | 18.79 | 21.40 | 3.09 | -18.18 |
| 12:00 PM | -21.13 | -2.19 | 18.84 | 21.54 | 3.07 | -18.44 |
| 15:00 PM | -21.1 | -2.13 | 18.87 | 21.52 | 3.03 | -18.46 |
| $max^\Delta|\delta|$ | 0.04 | 0.1 | 0.06 | 0.04 | 0.1 | 0.07 |
| $max^\Delta|\sin \delta|$ | 7E-4 | 2E-3 | 1E-3 | 6E-4 | 2E-3 | 1E-3 |
| $max^\Delta|\cos \delta|$ | 3E-4 | 7E-5 | 3E-4 | 3E-4 | 1E-4 | 4E-4 |

The mathematical expressions for different parameters $max^\Delta|\delta|$, $max^\Delta|\sin \delta|$ and $max^\Delta|\cos \delta|$ are given as follows:

$$max \Delta|\delta| = max(|\delta_9|,|\delta_{12}|,|\delta_{15}|) \quad (5)$$
$$- min(|\delta_9|,|\delta_{12}|,|\delta_{15}|);$$

$$max^\Delta|\sin \delta| \quad (6)$$
$$= max(|\sin \delta_9|,|\sin \delta_{12}|,|\sin \delta_{15}|)$$
$$- max(|\sin \delta_9|,|\sin \delta_{12}|,|\sin \delta_{15}|);$$

$$max^\Delta|\cos \delta| \quad (7)$$
$$= max(|\cos \delta_9|,|\cos \delta_{12}|,|\cos \delta_{15}|)$$
$$- max(|\cos \delta_9|,|\cos \delta_{12}|,|\cos \delta_{15}|);$$

Observing above calculated outcomes, it can be found that the change in the range of $\delta$, $\sin \delta$ and $\cos \delta$ are negligible and hence can be taken as constant. Hence, the solar declination at noon (i.e., 12:00 PM) can be stated as the signifying measure for the respective time instant or the date.

### 3.2 Solar hour angle

Typically, the inter-relation between different solar parameters such as solar elevation, solar azimuth and solar hour angle is expressed as follows:

$$\sin h_s = \sin \phi \sin \delta + \cos \phi \cos \delta \cos \Omega \quad (8)$$
$$\tan \alpha_S = \frac{\cos \delta \sin \Omega}{\cos h_s} \quad (9)$$
$$\cos \alpha_S = \frac{\sin h_s \sin \phi - \sin \delta}{\cos h_s \cos \phi} \quad (10)$$

where $h_s, \alpha_S, \phi, \delta,$ and $\Omega$ signify the solar elevation, solar azimuth; current latitude, solar declination and the solar hour angle, respectively.

On the basis of above derived equations, in our model, the solar hour angle has been estimated as follows (11):

$$\Omega = \begin{cases} min\begin{pmatrix} arccos\left(\frac{-b+\sqrt{b^2-4ac}}{2a}\right), \\ arccos\left(\frac{-b-\sqrt{b^2-4ac}}{2a}\right) \end{pmatrix} ; Morning \\ -min\begin{pmatrix} arccos\left(\frac{-b+\sqrt{b^2-4ac}}{2a}\right), \\ arccos\left(\frac{-b-\sqrt{b^2-4ac}}{2a}\right) \end{pmatrix} ; Afterno \end{cases} \quad (11)$$

where

$$a = tan^2 \alpha_S \sin^2 \phi + 1 \quad (12)$$
$$b = -\sin 2\phi \tan \delta \tan^2 \alpha_S \quad (13)$$
$$c = tan^2 \alpha_S \cos^2 \phi \tan^2 \delta - 1 \quad (14)$$

In this study, $\alpha_S$ and $\phi$ have been estimated using Google Earth GIRT tool, and the other parameter $\delta$ has been estimated using equation (2). On the basis of the initial parameters $\Omega$ of the remote sensing image under study has been estimated (11). Interestingly, here it can be observed that employing equation (11) the time stamp such as "Morning" or "Afternoon" can also be obtained by means of the direction of the shadow in the image under study.

### 3.3 Solar elevation

As in the previous phases, already the key parameters such as solar declination, solar hour angle and latitude are obtained. Thus using these parameters, the following equation (15) estimates the solar elevation($h_s$). Mathematically,

$$h_s = arcsin(\sin \phi \sin \delta + \cos \phi \cos \delta \cos \Omega) \quad (15)$$

## 4. MAN-MADE INFRASTRUCTURE HEIGHT ESTIMARION

This section discusses the dimensional characterization of the man-made infrastructures. In this research work, height estimation of a man-made structure has been performed. The discussion of the height estimation and neighbouring structures height assessment to ensure most accurate strategic decision is presented in following sub-sections.

### 3.2 Man-Made Structure Height Estimation

Typically, in generic perception, estimation of certain man-made (such as, building) or even natural structure is highly intricate and even seems infeasible using satellite image due to significantly huge distance. To deal with this issue, in this paper a method has been derived to estimate the height of a man-made structure, such as building. Here, height H is estimated using celestial geometry concept. Fig. 5 represents the inter-relation between shadow and a man-made construction, where image is taken from remote sensing satellite.

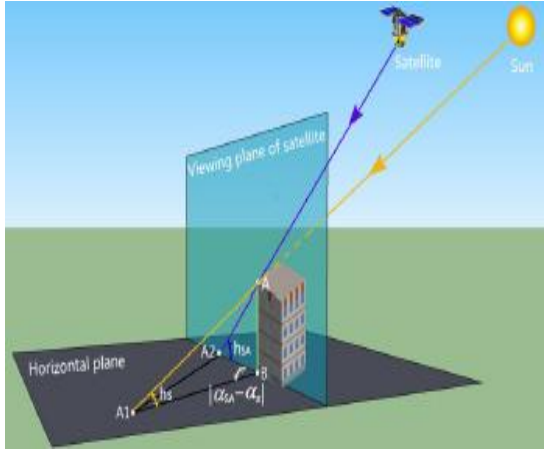

**Fig 4 Inter-relation between shadow length and man-made structural dimension (i.e., height)**

Observing above figure (Fig. 4) and considering triangle with coordinates (A, A1, B) and tangential relationship, the following equation can be derived to estimate length of the shadow:

$$L_{A1B} = \frac{H}{\tan h_S} \tag{16}$$

Similarly, from coordinates (A, A2, B),

$$L_{A2B} = \frac{H}{\tan h_{SA}} \tag{17}$$

Now, with the triangle coordinates (A1, A2, B), the cosine rule can be presented as:

$$L_{A1A2}^2 = L_{A1B}^2 + L_{A2B}^2 - 2L_{A1B}L_{A2B}\cos(\alpha_S - \alpha_{SA}) \tag{18}$$

Now, applying above equations (13, 14, and 15), the height of the man-made infrastructure can be obtained using following mathematical expression (16):

$$H = \tan h_S \left( L_{A2B} \cos(\alpha_S - \alpha_{SA}) + \sqrt{L_{A1A2}^2 - L_{A2B}^2 \sin^2(\alpha_S - \alpha_{SA})} \right) \tag{19}$$

Thus, applying the above approach, in this paper the height estimation of a man-made infrastructure has been done using shadow information and sun position information.

Considering proposed model to be used for certain strategic utilities such as identifying target locations, specific target dimension, etc (here we indicate utilities such as threat monitoring, supervision or even strategic offensive measures. Even though the proposed model can be used for building height estimation to assess energy consumption etc), ensuring precise identification of the structure (i.e., distinguishing a structure from others) is must. With this motivation, in this paper, the surrounding man-made structures have also been taken into consideration and examined for its height. A brief of the neighbouring structure assessment (height estimation) is discussed in following sub-section.

### 3.3 Neighbouring Structural Characterization

Referring the concept proposed and derived in [38], it can be stated that with remote sensing images, particularly retrieved through satellite, the ratio of structural height and its shadow length can be considered as constant [38]. Once estimating the height of a man-made structure or any infrastructure under supervision, the ratio of the infrastructure height to corresponding shadow length can be estimated as follows (20):

$$R_{cs} = \frac{H}{L_{A1A2}} \tag{20}$$
$$= \frac{\tan \alpha_S \left( L_{A2B} \cos \frac{(\alpha_S - \alpha_{SA}) +}{\sqrt{L_{A1A2}^2 - L_{A2B}^2 \sin^2(\alpha_S - \alpha_{SA})}} \right)}{L_{A1A2}}$$

To ensure optimal or most accurate target identification and respective structural characterization (say, height estimation), it can be suggested to measure the heights of the neighbouring structures and average them. This averaged height value can play a role of SIGNIFIER playing vital role to distinguish target from other structures. In this research and study the length acquisition tool of Google Earth application. Here, we have applied GIRT tool to measure shadow length of the other structures and eventually using $R_{cs}$ and the shadow length, the height of other structures have been obtained as (21):

$$H^* = R_{cs} \times L_{A1A2}^* \tag{21}$$

where, H* represents the height of neighbouring structures and $L_{A1A2}^*$ presents the corresponding shadow length.

In real time scenarios, there can be other entities around target or the structure under supervision such as tree, movable objects such as vehicles etc which can significantly impact on the efficiency. In addition, it can have impact on the accuracy of the shadow length. To deal with such limitations, in this paper an additional parameter $R_{HS}$ has been introduced. As already hypothesized that with remote sensing image or the satellite image, $R_{cs}$ can be considered as a constant value. Similarly, the supplementary parameter $R_{HS}$ can also be considered as constant and can be proved based on the similar deduction approach as applied for $R_{cs}$.

$$R_{cs} = \frac{H}{L_{A1A2}} \tag{22}$$
$$= \frac{\tan \alpha_S \left( L_{A2B} \cos \frac{(\alpha_S - \alpha_{SA}) +}{\sqrt{L_{A1A2}^2 - L_{A2B}^2 \sin^2(\alpha_S - \alpha_{SA})}} \right)}{L_{A1A2}}$$

Estimating vertical edge length $L_{A2B}$ using GIRT, the height of the neighbouring structures can be easily estimated even without proper ground conditions and under ambiguities. Mathematically, it can be obtained suing following mathematical expressions:

$$R_{HS} = \frac{H}{L_{A2B}} \tag{23}$$
$$H = R_{HS} L_{A2B} \tag{24}$$

Thus, applying proposed method, a low cost strategic infrastructure characterization, especially height estimation can be done.

To enable optimal performance using proposed method, ensuring defined relative error for the structure height is needed and on the basis of the defined relative error parameter, the maximum positive as well as negative slope can be estimated (1). In case the real slope is lower than the highest estimated slope, our proposed approach can be applied for estimating structural heights and neighbouring structures (heights) for any satellite image of the remote sensing image, fulfilling criteria such distinct vertical edge to enable accurate shadow length estimation and disambiguate or clear shadow presence. Being a well calibrated and optimistically derived model, our research approach enables dimensional characterization (height estimation) of a man-made structure even under low density structure (using (20) and (21)) as well as dense region such as urban reasons (using (23), (24)).

The results obtained and respective discussion is presented in the following section.

## 4. RESULTS AND DISCUSSION

Exploring the significance of geo-tagged information of a satellite image and associated shadow information in the image to enable man-made infrastructure dimensional characterization, in this paper a novel approach has been developed to estimate the height of a man-made structure accurately. To evaluate the performance and effectiveness of the proposed system in this paper few random satellite images fulfilling key criteria were taken into consideration.

The remote sensing satellite images were at first tested where it was assured that the image does have at least one distinctly vertical edge and crisp-clear shadow presence to enable accurate shadow length estimation. Considering seamless and optimistically threat less assessment, in this paper five man-made structural images have been selected randomly which have been obtained through Google Earth application. To ensure security aspects, in this manuscript the details of the infrastructures have not been disclosed. However the implementation model and its precise implementation might enable readers to perform expected structural height estimation using shadow information and geo-tagged image information such as solar parameters and time stamps on the satellite image. For a case study, one image sample image was taken into consideration.

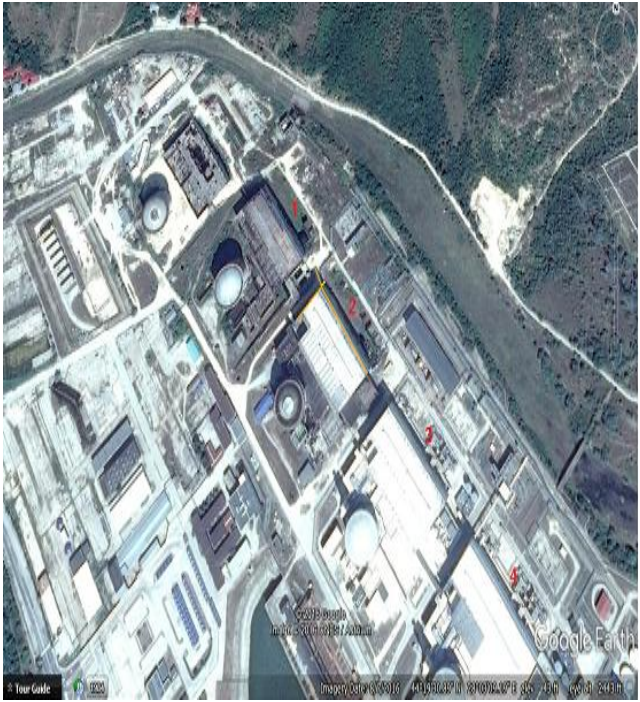

**Fig. 5 Sample Satellite image (Source: Radom image of a Nuclear power plant taken from Google Earth)**

A random building (Cairo, Egypt) was taken into consideration to perform its height estimation. We obtained the radio sensing image data directly using Google Earth GITS tool (here, length and angular acquisition tools). The details of the measurement process can be found in reference [38]. The measurements obtained are given in following table.

**TABLE 4. SATELLITE IMAGE INFORMATION**

| Date | RS Image Lat. | Solar Azimuth | Sat. Azimuth | $L_{A1A2}(m)$ | $L_{A2B}(m)$ |
|---|---|---|---|---|---|
| 08 Jan 2017 | $44.19^0$ | $142.69^0$ | $163.64^0$ | 3.20 | 9.49 |

**TABLE 5. TOWER HEIGHT**

| Solar declination | Solar hour angle | Solar elevation | Height of structure |
|---|---|---|---|
| $4.42^0$ | $-16.67^0$ | $58.92^0$ | 19.30 |

In this paper, to enable neighboring structure characterization and precise target identification $R_{cs}$ has been estimated for other neighboring structures. Here, the respective heights of the three neighboring structures have been estimated by means of proposed mathematical formulation discussed in previous sections.

The results obtained are as follows:

**TABLE 6. ESTIMATED STRUCTURAL HEIGHT BASED ON SHADOW INFORMATION**

| Structure No. | $L_{A2B}$ | $L_{A1A2}$ | $H$ | $H_m$ | Diff. | $R_{cs}$ |
|---|---|---|---|---|---|---|
| 1 | 9.15 | 3.26 | 19.20 | 18.98 | 0.22 | 5.53 |
| 2 | 9.25 | 3.24 | 18.13 | 17.51 | 0.62 | 5.74 |
| 3 | 9.45 | 3.49 | 19.24 | 18.15 | 1.09 | 5.54 |
| 4 | 9.24 | 3.28 | 18.45 | 18.21 | 0.24 | 5.63 |
| Avg-5.61, Std. dev.-0.097, Time of Data Acquisition: 01:40 PM | | | | | | |

Based on the retrieved heights of the man-made construction, the average $R_{cs}$ of the building was obtained as 5.61 meter and the standard deviation was found as 0.097. Thus, in the presented study, the $R_{cs}$ representative index can be stated to be 5.61 for the considered satellite image. Similarly, using equation (21) the height of the other neighboring structures can be obtained. To further evaluate the performance of the proposed dimensional (i.e., height) estimation model, in this paper some other existing research outcomes have been compared with the proposed model. We have performed performance assessment in terms of means error (meter). Here, we have applied standard deviation of structural heights as the signifier for mean error. The accuracy analysis or comparison of the proposed model is presented in Table 7.

TABLE 7 Performance Comparison

| Techniques | Height Range | Error (meter) |
|---|---|---|
| [31] | NA | 1.88 |
| [35] | 1.96-18.87 | 0.87 |
| [47] | NA | 2.80 |
| [48] | 32-168 | 12.99 |
| Proposed | 17.51-18.98 | 0.97 |

As depicted in above mentioned comparative performance outcomes, it can be found that the proposed method exhibits better accuracy than major existing approaches. Here, it can be seen that with higher height range, the estimation turns out giving error prone results or inaccuracy increases with higher height structures. These experimental results also revealed that the accuracy of the estimation can be higher only with the structures having slope lower than 30 degree. Unlike major traditional approaches, the proposed model avoids inevitable needs of information such as ground object prior information, sun/satellite-parameters, shadow information etc to estimate man-made structural height. It can be of great significance for monitoring and surveillance purposes particularly for UAV based strategic applications.

## 6. CONCLUSION

In this paper, a multi-tech model comprising shadow detection and information retrieval, Geo-tagged information extraction have been applied to estimate the height of the strategic man-made constructions. The overall proposed model comprises, satellite image retrieval, shadow detection and shadow length estimation, geo-tagged information retrieval using Geometry and Information Retrieval Tool Google Earth tool and target identification, its height estimation and neighbouring structural height estimation. Unlike major existing approaches, being an outdoor satellite image analysis function, in this work Google Earth Geometry and Information Retrieval Tool has been applied to extract significant solar and shadow information, such as latitude, solar azimuth, and solar elevation. In addition, this research work examined and employed significant mathematical models to estimate the solar declination, solar hour angle, etc that plays vital role in estimating the height of the man-made infrastructure using satellite image. However, the proposed approach can be applied with the remote sensing satellite images having crisp clear edge information and precise shadow length (shadow presence) availability. The results state that the proposed method can be effectively used when the man-made structure height is expected to be less than 20 meters and slope remains less than 3 degree. The overall man-made structure height estimation result affirms that the proposed method outperforms other techniques. Being a free available tools and source based system, it can be used in major civil as well as defence purposes to monitor, identify and classify man-made structures from different neighbouring and ground information to make optimal decision process. Since, the proposed system greatly depends on the shadow length and associated dimensions to estimate other solar parameters, and therefore in future certain efficient shadow detection and length estimation model can be used to enhance the proposed model.

# Columns on Last Page Should Be Made As Close As Possible to Equal Length

# Authors' background

| Your Name | Title* | Research Field | Personal website |
|---|---|---|---|
| Romany F. Mansour | assistant professor | Computer Science | |
| Elsaid MD. Abdelrahim | assistant professor | Computer Science | |
| Abdulsamad Al-Marghilnai | associate professor | Computer Science | |

**\*This form helps us to understand your paper better, the form itself will not be published.**

**\*Title can be chosen from: master student, Phd candidate, assistant professor, lecturer, senior lecture, associate professor, full professor**